\documentclass[12pt]{article}
\usepackage{graphicx}
\usepackage{amssymb}
\usepackage{amsmath}

\newcommand{\tb}{\tan{\beta}}
\newcommand{\tbb}{\tan^{2}{\beta}}
\newcommand{\tbbb}{\tan^{3}{\beta}}
\newcommand{\sa}{\sin{\alpha}}
\newcommand{\ca}{\cos{\alpha}}
\renewcommand{\sb}{\sin{\beta}}
\newcommand{\cb}{\cos{\beta}}
\newcommand{\sab}{\sin{(\alpha - \beta)}}
\newcommand{\cab}{\cos{(\alpha - \beta)}}

\newcommand{\GeV}{ {\rm GeV} }
\newcommand{\TeV}{ {\rm TeV} }

\newcommand{\BR}{\text{BR}}
\newcommand{\meg}{\mu \to e \gamma}

\newcommand{\mnen}{\mu {\rm N} \to e {\rm N}}
\newcommand{\maleal}{\mu {\rm Al} \to e {\rm Al}}
\newcommand{\bsmumu}{{B}_s\rightarrow\mu^{+} \mu^{-}}
\newcommand{\doubleratio}{\BR(\mnen)/\BR(\meg)}

\newcommand{\MSUSY}{M_{\text{SUSY}}}

\begin{document}

\begin{titlepage}

\begin{flushright}
\end{flushright}
\vskip 1.35cm

\begin{center}

{\large Mass Spectrum Dependence of Higgs-mediated $\mu$-$e$ Transition in the MSSM}

\vskip 1.2cm

\renewcommand{\thefootnote}{\fnsymbol{footnote}}

Masaki Jung Soo Yang$^{a,b}$

\renewcommand{\thefootnote}{\arabic{footnote}}

\vskip 0.4cm

{\it $^a$Department of Physics, University of Tokyo, Tokyo 113-0033,
 Japan }\\
{\it $^b$Department of Physics, Nagoya University, Nagoya 464-8602, Japan}\\

\date{\today}

\begin{abstract} 

In this paper, we study non-decoupling $\mu$ - $e$ transition effects by Higgs-mediated contribution in the MSSM, when some SUSY mass parameters are much greater than the TeV scale. In order to treat CP-odd Higgs mass $m_{A^{0}}$ as a free parameter, we consider the non-universal Higgs mass model (NUHM), and assume the only left- or right-handed sleptons had flavor-mixing mass terms. 
As a result, when only right-handed sleptons have flavor-violation, there is some Higgs-dominant region although SUSY particle masses are around the TeV scale. We found it is necessary to consider the Higgs effect in the region where the gaugino effect receives destructive interference.
Moreover, the ratio of branching ratios $\BR(\meg) / \BR(\maleal)$ drastically depends on the mass spectrum structure and chirality of flavor violation.
The log factor from two split mass scales influences the way of interference between gaugino- and Higgs-mediated contributions significantly.

\end{abstract} 



\end{center}
\end{titlepage}

\section{Introduction}

The charged lepton-flavor violating (cLFV) processes are 
the clear probe for the physics beyond the standard model (SM) \cite{Raidal:2008jk}.
Finite neutrino masses introduce these cLFV processes, nevertheless,  
they are too tiny to observe. Therefore, if we found any positive signal of such processes, it would be reliable evidence of new physics. 

Now the MEG experiment is searching for $\meg$
\cite{Adam:2009ci}. It aims at a sensitivity of $\sim 10^{-13}$ for the branching ratio
on the first stage, which is two orders of magnitude
below the current bound. The COMET and Mu2e experiments
\cite{mu2e,comet}, which are searches for $\mu$-$e$ conversion in nuclei ($\mnen$), are being planned in J-PARC and Fermilab, respectively. It is
argued that they would reach  $\sim 10^{-16}$ for the branching ratio of
$\mu$-$e$ conversion with target $\rm Al$. 

In the supersymmetric (SUSY) framework, cLFV is induced by misalignment of the mass matrix between the fermion and the sfermion, originating from SUSY-breaking terms (we call these effects gaugino-mediated contribution).
Moreover it is well known that a non-holomorphic correction by the SUSY-breaking effect induces 
additional flavor violation in the Higgs sector \cite{Babu:1999hn,Babu:2002et} 
(we call these effects Higgs-mediated contribution).

Branching ratios for the cLFV processes due to the gaugino-mediated contribution are suppressed by a typical mass scale of the SUSY particles $\MSUSY$, since the effective dipole interaction is dominant in the cLFV processes. 
On the other hand, since the Higgs-mediated effects are not suppressed by $\MSUSY$ \cite{Babu:2002et},
these contributions to the processes could be sizable when $\MSUSY$ is much greater than the TeV scale. 
The branching ratio of
$\mu$-$e$ conversion in nuclei is more sensitive to the Higgs-mediated
contribution \cite{Kitano:2002mt,Kitano:2003wn}. Indeed, the ratio of the branching
ratio for $\mu\rightarrow e \gamma$ and $\mu$-$e$ conversion in
nuclei is a good observable to constrain mass spectrum in the MSSM,
since it is sensitive to whether the gaugino-mediated or
Higgs-mediated contribution is dominant.

These non-holomorphic cLFV effects have been much studied in the literature \cite{Kitano:2002mt, Kitano:2003wn, Dedes:2002rh, Black:2002wh, Brignole:2003iv}, and
it was pointed out that two-loop Barr--Zee diagrams become dominant in the Higgs-mediated contribution \cite{ Paradisi:2005tk, Paradisi:2006jp}. However, a
systematic calculation including these Barr--Zee diagrams has not been performed.

In a previous study \cite{Hisano:2010es}, we discussed these Higgs-mediated $\mu$ - $e$ transition effects in the MSSM, including dominant Barr--Zee diagram contributions.
It was assumed that only the left-handed sleptons had flavor-mixing terms. 
In the case in which all SUSY particles are degenerate, with a common 
mass $\MSUSY$ which is much greater than the TeV scale, we have discovered that Higgs- and gaugino-mediated 
dipole amplitudes became comparable to each other when $\MSUSY/m_{A^0} \sim 50$.
However, these assumptions are not always robust.
This common mass approximation is too crude, and SUSY particles might be large splitting  spectrum.
Furthermore, if right-handed sleptons have flavor-mixing terms,  
the gaugino-mediated contribution receives destructive interference 
between the bino and bino-Higgsino amplitudes \cite{Hisano:1996qq}. 
This cancellation structure could be modified in the Higgs dominant region. 

 Therefore, in this study, we promote the previous analysis in a more precise form. 
We research these non-decoupling cLFV effects, in some SUSY mass spectra and flavor-violation sources of the MSSM.
In order to treat CP-odd Higgs mass $m_{A^{0}}$ as a free parameter, we relax universality of scalar soft masses $m_{0}$ for the MSSM Higgs multiplets. This model is called the non-universal Higgs mass model (NUHM), which is a generalization of the minimal supergravity (mSUGRA) model. 
As flavor-violation sources, we assume only the left- or right-handed sleptons have flavor-mixing mass terms. These setups are realized in the SUSY seesaw and pure SUSY $SU(5)$ grand unified theory (GUT) models,  respectively \cite{Borzumati:1986qx,Barbieri:1994pv}. We do not treat Higgs-mediated LFV effects from the left-right sleptons mixing because it is suppressed by a factor $v/\MSUSY$, by consequence of gauge invariance.
If both Higgs-mediated and ordinary SUSY contribution are significant, the ratio of branching ratios $\BR(\meg) / \BR(\maleal)$ becomes sensitive to SUSY mass parameters. We investigated these mass-sensitive regions and the behavior of the ratio $\BR(\meg) / \BR(\maleal)$ in some mass spectrum of the NUHM.

As a result, when only right-handed sleptons have flavor-violation, there is some Higgs-dominant region although SUSY particle masses are around the TeV scale. We found it is necessary to consider the Higgs effect in the region where the gaugino effect receives destructive interference.
Moreover, when sleptons are much heavier than charginos, cf. split SUSY models \cite{ArkaniHamed:2004fb},  
we found that the ratio $\BR(\meg) / \BR(\maleal)$ drastically depends on the mass spectrum structure and chirality of flavor violation. 
The log factor from two split mass scales influences the way of interference between gaugino- and Higgs-mediated contributions significantly.

This paper is organized as follows. In the next section we review the
Higgs- and gaugino-mediated contributions and formulae for
$\BR(\meg)$ and $\BR(\mnen)$. In Sec.~3, we discuss competition of the 
Higgs- and gaugino-mediated contributions in the NUHM, and 
how these effects influence physical observable $\doubleratio$   
when some SUSY mass parameters are much greater than the TeV scale.
Sec.~4 is devoted to conclusions and discussion.

\section{Lepton-Flavor Violation in the MSSM}

In this section, we present a theoretical framework and 
analytic formulae of LFV amplitudes and branching ratios.
The MSSM superpotential is given by
\begin{equation}
W = Y^{u}_{ij} U^{c}_{i}  Q_{j} H_{2} -  Y^{d}_{ij} D_{i}^{c} Q_{j} H_{1} - Y^{e}_{ij} E_{i}^{c} L_{j} H_{1} + \mu H_{2} H_{1} ,
\end{equation}
where Yukawa couplings $Y^{u}_{ij}$, $Y^{d}_{ij}$, and $Y^{e}_{ij}$ are
 3 by 3 matrices in family space, and $\mu$ is the Higgsino mass.
LFV interactions in the Higgs sector are generated by 
a non-holomorphic correction by a SUSY-breaking effect.
In the mass-eigenstate basis for both leptons and Higgs 
bosons, one leads to the effective Lagrangian \cite{Babu:2002et}
\begin{equation} 
\begin{split}
   \mathcal{-L}_{\rm eff}^{i \neq j} 
   &= \frac{(8 G_F^2)^{\frac{1}{4}} }{ \cos^2{\beta}} 
   \overline{e}_{R i} ( m_i \Delta^{L}_{ij} 
   + \Delta^{R}_{ij} m_j )  e_{L j}   
   \times \frac{1}{\sqrt{2}} 
   (h^0 \cab + H^0 \sab - i A^0 ) \\
   &~~+ \frac{(8 G_F^2)^{\frac{1}{4}} }{ \cos^2{\beta}} 
   \overline{e}_{R i} ( m_i \Delta^{L}_{ij} + \Delta^{R}_{ij} m_j )  
   \nu_{e L j}  H^- + {\rm h.c.}, 
\end{split}
\label{LFV}
\end{equation}
where $G_{F}$ is the Fermi constant, $m_{i,j}$ are the charged lepton masses, $\alpha$ is the mixing angle between the CP-even Higgs bosons $h_0$ and 
$H_0$, $A_0$ and $H^{\pm}$ are CP-odd and charged Higgs boson respectively, and $\tan\beta$ is defined by $\tan\beta = v_{u}/v_{d}$.

At the one-loop level, these $\Delta^{L,R}$ are
induced by the exchange of gauginos / Higgsinos and sleptons, 
and flavor-violation slepton masses.
In the mass insertion approximation, these $\Delta^{L,R}$ are 
induced by three diagrams, $SU(2)$ wino-Higgsino exchange, 
$U(1)$ pure bino exchange, and $U(1)$ bino-Higgsino one,  
expressed as follows \cite{Babu:2002et, Paradisi:2005tk, Paradisi:2006jp}:
\begin{equation}
\begin{split}
   \Delta^{L}_{ij} 
   &= - \frac{\alpha_{1}}{4\pi}\mu M_1 \delta^{LL}_{ij} m_{L}^2
   \left[  I_{4} (M_1^2, m_{R i}^2, m_{L i}^2,m_{L j}^2)+\frac{1}{2} I_{4} 
   (M_1^2, \mu^2, m_{L i}^2, m_{L j}^2)   \right]   \\
   &~~+ \frac{3}{2} \frac{\alpha_{2}}{4\pi} \mu M_2 \delta^{LL}_{ij} 
   m_{L}^2 I_{4} (M_2^2, \mu^2, m_{L i}^2, m_{L j}^2) ~,  \\
   \Delta^{R}_{ij} 
   &= \frac{\alpha_{1}}{4\pi}\mu M_1 \delta^{RR}_{ij} m^{2}_{R}
   \left[ I_{4} (M^{2}_{1},\mu^2,m^{2}_{R i} , m^{2}_{R j})\!-\! 
   (\mu\! \leftrightarrow\! m_{L i}) \right] .
\end{split}
\label{Delta}
\end{equation}
Here $\alpha_{1,2}$ are $g_{1,2}^{2} / 
4\pi$, $M_{1,2},~  m_{(L,R) i}$ are the gaugino and slepton masses, 
 $\delta^{(LL ,RR)}_{ij}$ are the mass insertion 
parameters, $\delta^{LL (RR)}_{ij} = (
m^2_{\tilde{l}_L (\tilde{e}_{R})})_{ij}/{m}^2_{L (R)}$, where $ (
m^2_{\tilde{l}_L (\tilde{e}_{R})})_{ij}$ are off-diagonal elements of left- (right-) handed
sleptons mass matrix and ${m}_{L (R)}$ is an average
left- (right-) handed slepton mass.
The loop function $I_{4}(a,b,c,d)$ is given by 
\begin{equation}
 I_{4}(a,b,c,d) = - \left\{ { a \log a \over (a-b)(a-c)(a-d) } + {\rm cyclic} \right\} .
\end{equation}
Both $\Delta^{L}$ and $\Delta^{R}$ couplings suffer from strong cancellations due to destructive interference among various diagrams \cite{Brignole:2003iv}. In particular, in the $\Delta^{R}$ case, $\Delta^{R}$ vanishes when $\mu = m_{L_{i}}$.

\subsection{$\meg$ and dipole amplitudes}
Now we discuss about the LFV processes, at first, $\mu \to e \gamma$ process in the SUSY framework.
The effective amplitude of lepton radiative decay $l_{i}^{+} \to l_{j}^{+} \gamma$ is parametrized as 
\begin{equation}   \label{T}
   T = e \epsilon^{\mu *}(q) ~ \overline{v}_i (p) 
   \biggl[ m_{l_i} i \sigma_{\mu \nu} q^{\nu} 
   (A^{L}_{ij} P_{L} + A^{R}_{ij} P_{R}) \biggr] v_{j} (p-q).
\end{equation}    
In this study, we mainly consider the gaugino- and Higgs-mediated amplitudes for the $\mu$ - $e$ transition, where the flavor indices are taken to be $(i,j) = (\mu , e)$.
The branching ratio of $\meg$ is written as 
\begin{equation}
\BR(\meg) = { 48 \pi^{3} \alpha_{\rm em} \over G_{F}^{2}} (|A^{L}_{\rm gaugino}+A^{L}_{\rm Higgs}|^{2} + |A^{R}_{\rm gaugino}+A^{R}_{\rm Higgs}|^{2}) ,
\end{equation}
where $\alpha_{\rm em}$ is the fine structure constant, and $A^{(L,R)}_{\rm (gaugino,Higgs)}$ are gaugino- / Higgs-dominated dipole amplitudes for each chirality.

In the mass insertion approximation, the effective amplitudes of the gaugino-mediated LFV proportional to $\tb$ are given by \cite{Hisano:1995cp, Hisano:1998fj}
\begin{equation}
\begin{split}
A^{L}_{\rm gaugino} &\simeq {\alpha_{2}\over4\pi} \mu M_{2} \delta^{LL}_{ij} m_{L}^2 \tb \\
&\times D \left[ D\left[ {1\over m^{2}} \left\{ - f_{c} ({M^{2}\over m^{2}}) + {1\over4} f_{n}({M^{2}\over m^{2}}) \right\} ; M^{2}  \right] (M_{2}^{2},\mu^{2}) ; m^{2} \right] (m_{L i}^{2},m_{L j}^{2}) ,\\
\\
A^{R}_{\rm gaugino} &\simeq -{\alpha_{1}\over4\pi} \mu M_{1} \delta^{RR}_{ij} m_{R}^2 \tb \\
&\times D \left[ D\left[ {1\over2 m^{2}} f_{n} ({M_{1}^{2} \over m^{2}}) ; m^{2}  \right] (m_{R}^{'2},m_{L i}^{2}) ; m^{'2}_{R} \right] (m_{R i}^{2},m_{R j}^{2}) \\
&+ {\alpha_{1}\over4\pi} \mu M_{1} \delta^{RR}_{ij} m_{R}^2 \tb \\
&\times D \left[ D\left[ {1\over2 m^{2}} f_{n} ({M^{2} \over m^{2}}) ; M^{2}  \right] (M_{1}^{2}, \mu^{2}) ; m^{2} \right] (m_{R i}^{2},m_{R j}^{2}) , \\
\end{split}
\end{equation}
where $D[f(x);x] (x_{1},x_{2}) = (f(x_{1}) - f(x_{2})) / (x_{1} - x_{2}),$ and
\begin{equation}
\begin{split}
f_{c} (x) &= - {1\over 2(1-x)^{3}} (3 - 4x + x^{2} + 2 \ln x) \ , \\
f_{n} (x) &= {1\over(1-x)^{3} } (1 - x^{2} + 2x \ln x) \ .
\end{split}
\label{loopint}
\end{equation}
In this analysis, we do not consider double insertion (such as $\delta^{LL}_{\mu \tau} \delta^{LL}_{\tau e} $ contributions) effects for simplicity.

\begin{figure}[h]
\begin{center}
\begin{tabular}{ccc}
   \includegraphics[width=7cm,clip]{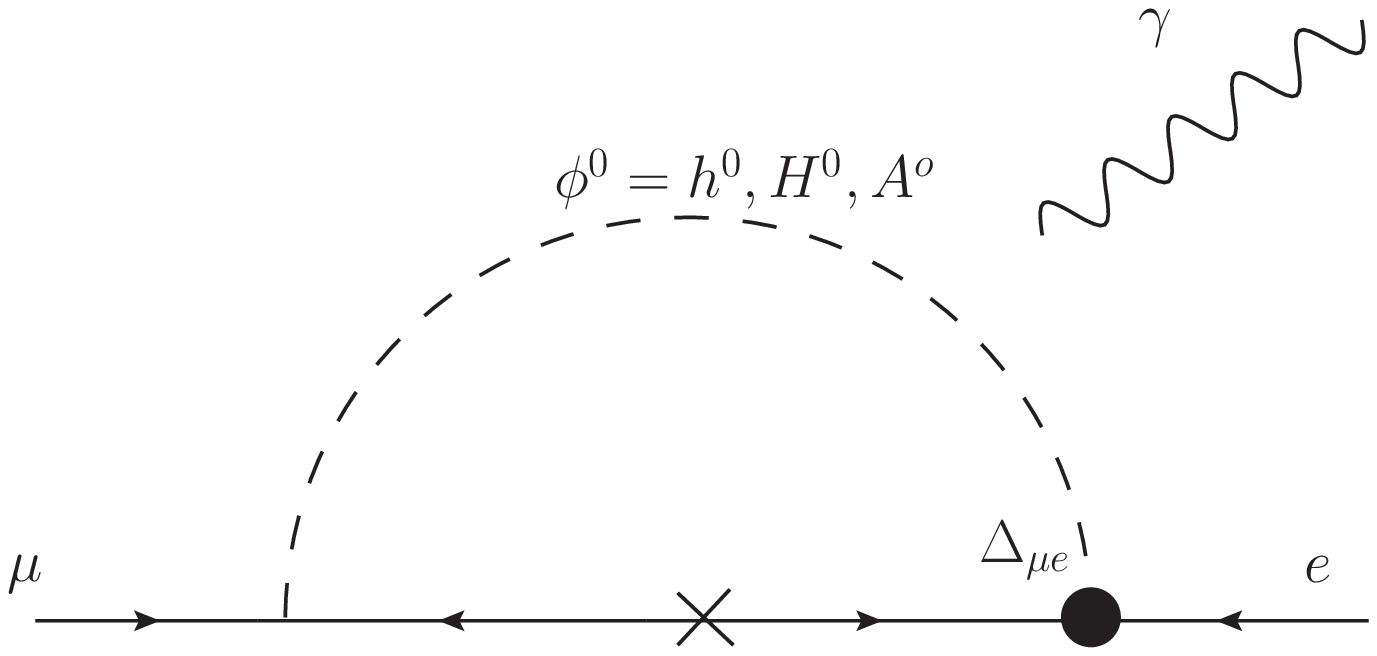} &
   \includegraphics[width=7cm,clip]{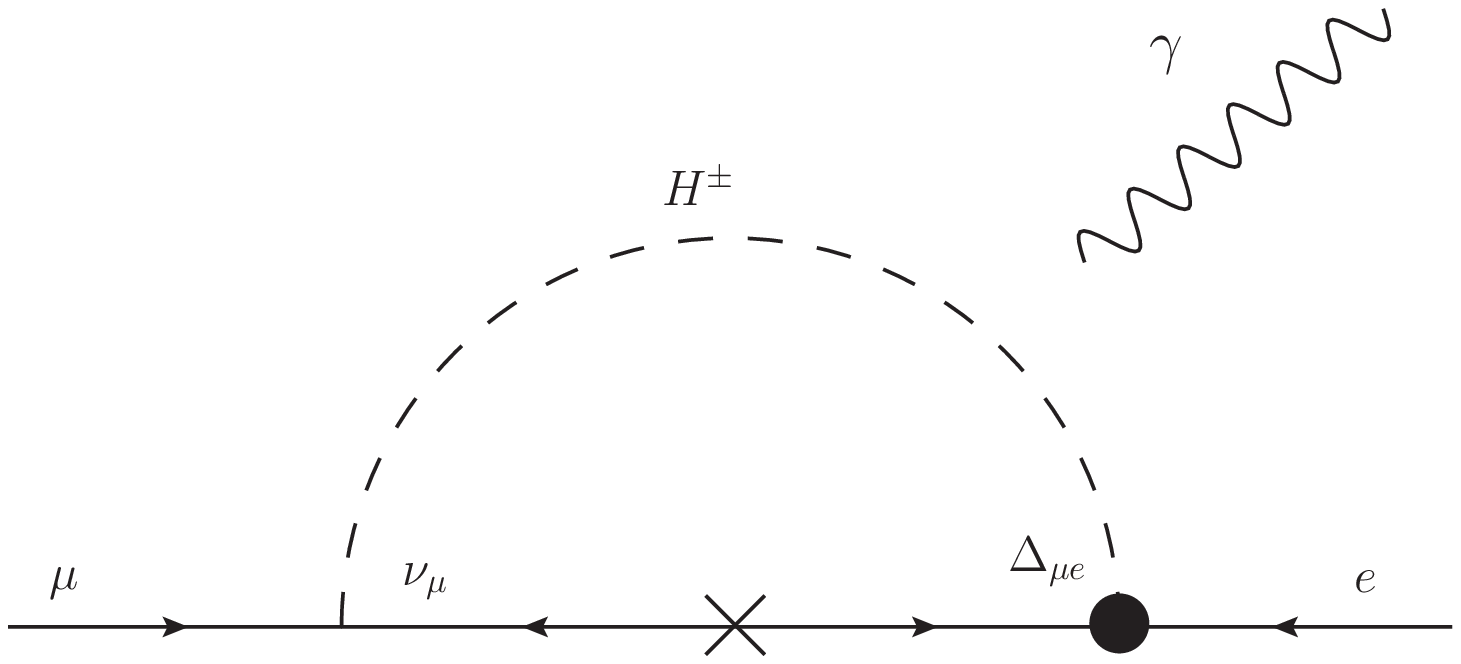} \\
   (a) & (b) 
\end{tabular} 
\caption{$\mu\rightarrow e \gamma$ induced by Higgs boson exchange at one-loop level.}
\label{1loop}
\end{center}
\end{figure}

Next we consider the Higgs-mediated contribution of dipole amplitudes. 
While these amplitudes could be induced at the one-loop level (Fig.~\ref{1loop}), it is
suppressed by three chiral flips, {\it i.e.}, one chirality flip in
the lepton propagator and two lepton Yukawa couplings. Indeed two-loop
diagrams may make a significant contribution. As shown in Fig.~\ref{Barr
  Zee}, two-loop diagrams, called Barr--Zee diagrams, involve only
one chiral flip (from lepton Yukawa coupling), and hence their
contribution is much larger than that at the one-loop level.

\begin{figure}[h]
\begin{center}
\begin{tabular}{ccc}
   \includegraphics[width=7cm,clip]{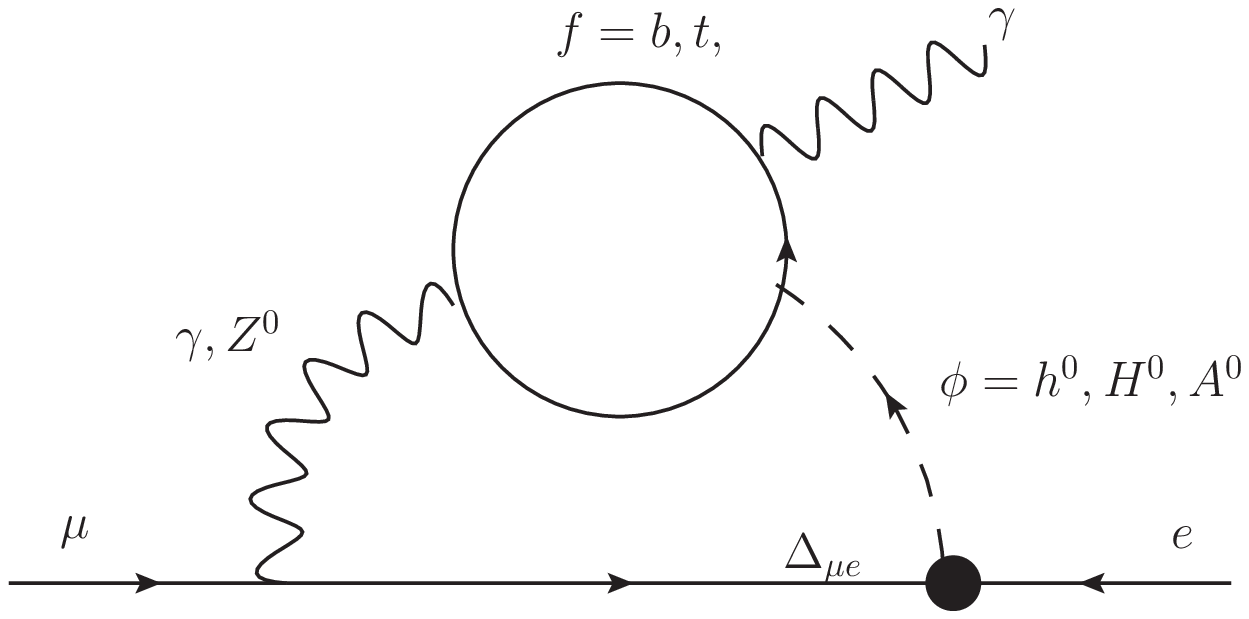} &
   \includegraphics[width=7cm,clip]{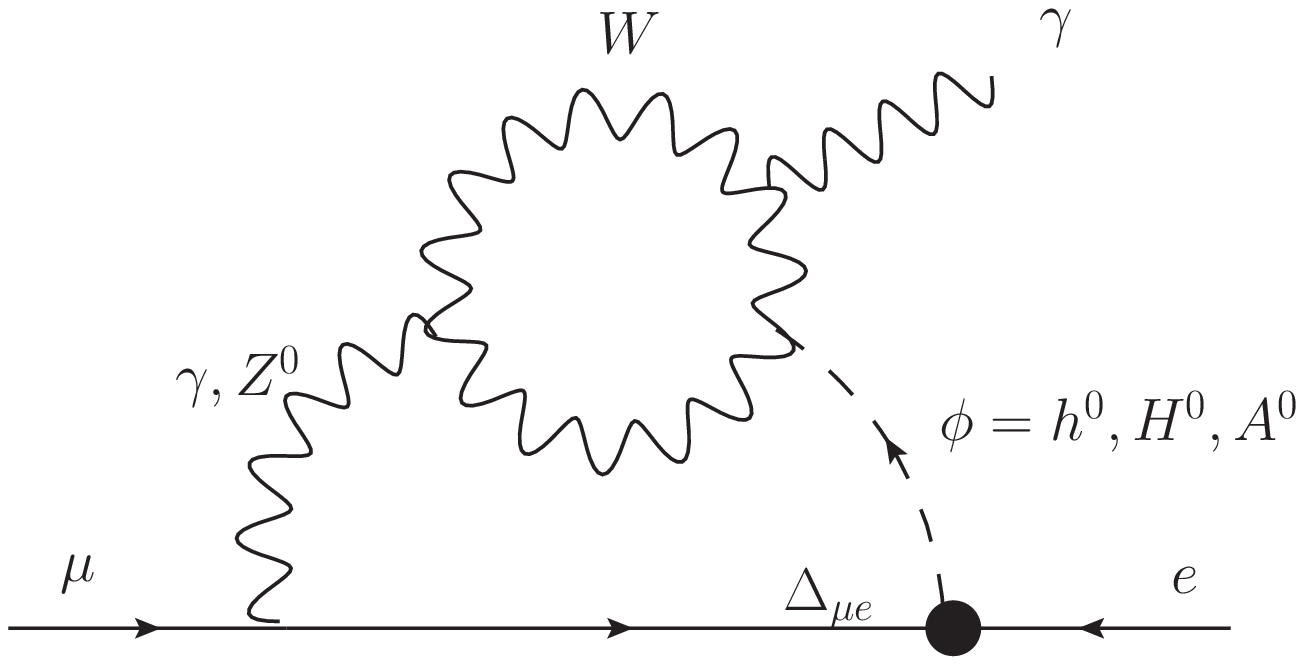} \\
   (a) & (b) 
\end{tabular} 
\caption{Examples of two-loop Barr--Zee diagrams induced by Higgs exchange.}
\label{Barr Zee}
\end{center}
\end{figure}

Following Refs.~\cite{Paradisi:2005tk,Paradisi:2006jp}, we consider
Barr--Zee diagrams that involve effective
$\gamma$-$\gamma$-$\phi^0$ vertices ($\phi^0=h^0$, $H^0$, and $A^0$). The
effective vertices are induced by heavy fermion or weak gauge/Higgs boson
loops.  It is found that Barr--Zee diagrams including
the top-quark loop (Fig.~\ref{Barr Zee}~(a)) give
a dominant contribution to $\mu\rightarrow e\gamma$, and the
bottom-quark one (Fig.~\ref{Barr Zee}~(a)) is also sizable 
only when $\tan\beta$ is large. The $W^-$
and NG-boson diagrams (Fig.~\ref{Barr Zee}~(b)) 
tend to be subdominant unless $m_{A^0}$ is small \cite{Hisano:2010es}.

These Barr--Zee diagrams involving fermion loops
 give contributions to the coefficient $A^{L,R}$
in Eq.~(\ref{T})  as
\begin{equation}
\begin{split}
   A^{(L,R)}_{{\rm BZ}(b)}  
   &=  {2 \sqrt{2}G_{F}  \alpha_{\rm em} N_c Q_{b}^{2} \over 16\pi^{3} } 
   \Delta^{(L,R*)}  \\
   &\times \left[ - {\cab \sa \over \cos^{3}\beta }  
   f(z_{h^0}^{b}) + {\sab \ca \over \cos^{3}\beta }  f(z_{H^0}^{b}) 
   + {\sb \over \cos^{3} \beta}  g(z_{A^0}^{b}) \right] , \\
   A^{(L,R)}_{{\rm BZ}(t)}  
   &=  {2 \sqrt{2}G_{F} \alpha_{\rm em} N_c Q_{t}^{2} \over 16\pi^{3} } 
   \Delta^{(L,R*)}  \\
   &\times \left[ {\cab \ca \over \cos^{2}\beta \sb}  
   f(z_{h^0}^{t}) +  {\sab \sa \over \cos^{2}\beta \sb} f(z_{H^0}^{t}) 
   + {1 \over \sb \cb} g(z_{A^0}^{t}) \right] . \\
\end{split}   \label{Abt}
\end{equation}
Here, $N_c =3$ is a color factor, and $Q_{b(t)}$ represents the electric
charge for the bottom (top) quark,  $\Delta^{(L,R*)}$ is $\Delta^{L}_{\mu e}$ for $A_{L}$, 
and $\Delta^{R*}_{e \mu}$ for $A_{R}$. $z_{\phi^0}^{q} =
m_{q}^{2}/m_{\phi^0}^{2}$ for $\phi^0 = h^{0}$, $H^{0}$, $A^{0}$ and
$q = b$, $t$.  Similarly, we calculate the coefficient for the tau-lepton
loop by substituting $N_{c} =1$, and replacing $Q_{b}, m_{b}$ with
$Q_{\tau}, m_{\tau}$.  The functions $f(z)$ and $g(z)$ are called
Barr--Zee integrals, given by
\begin{equation}
\begin{split}
   f(z) 
   &= {1 \over 2} z \int_{0}^{1} dx 
   { 1-2x(1-x) \over x(1-x) -z } \log{x(1-x) \over z} \ ,\\
   g(z) 
   &=  {1 \over 2} z \int_{0}^{1} dx 
   { 1 \over x(1-x) -z } \log{x(1-x) \over z} \ .
\end{split}   
\end{equation}
In the limit of $1 \gg z$, the asymptotic forms are given as follows \cite{Chang:1993kw}:
\begin{equation}
   f(z) \sim {z\over 2} (\log z)^{2} \ ,~   
   g(z) \sim {z\over 2} (\log z)^{2} \ .
 \end{equation}
On the other hand, in the limit of $1 \ll z$, 
\begin{equation}
  f(z) \sim {1 \over 3} \log z+ {13 \over 18} ,~  
   g(z) \sim {1 \over 2} \log z+ 1 . 
 \end{equation}

For $m_{A^0}\gg m_{Z^0}$ and $\tan \beta \gg 1$,
the Higgs-mediated contribution is approximated as
\begin{equation}
\begin{split}
   A_{{\rm Higgs}}^{(L,R)} 
   & \simeq {\sqrt{2} G_{F} N_{c} \alpha_{\rm em} \over 8 \pi^{3}} 
   \Delta^{(L,R*)}  \\
   &~~\times 
   \left[ 
   { Q_t^2 m_t^2 \over m_{A^0}^2 } 
   \tb \left(\log{m_t^2 \over m_{A^0}^2 } \right)^2
   - { Q_b^2 m_b^2 \over m_{A^0}^2 } \tan^3 \beta 
   \left( \log{m_b^2 \over m_{A^0}^2 } + 2\right) 
   - (b \rightarrow \tau) \right] \ . 
\end{split}
\label{ampHiggs}
\end{equation}
It is found that $\tan\beta$ and/or large logarithmic factors enhance
heavy-Higgs ($H^{0}$, $A^{0}$) contributions, and the light-Higgs
($h^{0}$) contribution is subdominant.

\subsection{$\mu$-$e$ conversion in nuclei}
Next let us discuss the $\mnen$ process. The conversion rate is formally given by \cite{Cirigliano:2009bz},
\begin{equation}
\begin{split}
\Gamma_{\rm conv} &\simeq {m_{\mu}^{5} \over 4} | C_{DL} D + 4G_{F} m_{\mu} (m_{p} \tilde{C}^{(p)}_{SL} S^{(p)} + m_{n} \tilde{C}^{(n)}_{SL} S^{(n)} ) |^{2} \\
&+ \, {m_{\mu}^{5} \over 4} | C_{DR} D + 4G_{F} m_{\mu} (m_{p} \tilde{C}^{(p)}_{SR} S^{(p)} + m_{n} \tilde{C}^{(n)}_{SR} S^{(n)} ) |^{2},
\end{split}
\end{equation}
where $D$, $S^{(N)}$ ($N = p,n$) are the dimensionless integrals of the overlap of the electron and the muon wave functions, $D=0.0360, S^{(p)} = 0.0153$, and $S^{(n)} = 0.0163$ for aluminium respectively. 
 $\tilde{C}^{(N)}_{S(L,R)}$ are nucleon-level effective scalar coefficients that contain QCD information, 
\begin{equation}
\tilde{C}^{(p,n)}_{S(L,R)} =  \sum_{q=u,d,s} C^{(q)}_{S(L,R)} \; f^{(q)}_{S(p,n)} 
\ + \ C_{GQ(L,R)} \; (1-\sum_{q=u,d,s} f^{(q)}_{S(p,n)}) ,
\end{equation}
where $C^{(q)}_{S(L,R)}$ are quark-level effective scalar coefficients, $f^{(q)}_{S(p,n)}$ are nucleon form factors for each quark $q$, $C_{GQ(L,R)}$ are effective coefficients of lepton-gluon vertex $e P_{(L,R)} \mu G^{\rho \nu}_{a} G^{a}_{\rho \nu}$, and detailed notations are given in Ref.~\cite{Cirigliano:2009bz}.

The dominant contribution to Higgs-mediated $\mnen$ comes from a tree-level Higgs boson exchange
\cite{Kitano:2003wn}. The reason is that coupling between the Higgs
boson and nucleon is characterized by the nucleon mass $m_N$ through
the conformal anomaly relation~\cite{Shifman:1978zn}, and could evade 
suppression of the light constituent quark mass. In the large $\tb$ region, down-type quarks receive 
an additional $\tb$ factor compared to  up-type quarks, and effective scalar coefficients are given by
\begin{equation}
 \tilde{C}_{S(L,R)}^{(N)} =  {\Delta^{(L,R*)}_{\mu e} \over m_{H^{0}}^{2} } (f_{S(N)}^{(d)} + f_{S(N)}^{(s)} + {2\over27} (1-\sum_{q=u,d,s} f^{(q)}_{S(N)}) ) \tbbb,
\end{equation}
where $m_{H^{0}}$ is the heavy CP-even Higgs mass, and other objects are the same as before. The last term comes from the bottom-quark, which is integrated out in the low energy effective theory. We neglect monopole and box diagrams in large $\tb$.
Dipole amplitudes in our convention are expressed as follows:
\begin{equation}
 C_{D(L,R)} =  {1\over2} e ( A^{(L,R)}_{\rm gaugino} + A^{(L,R)}_{\rm Higgs} ),
\end{equation}
where $e$ is the electric charge. The branching ratio of $\mu \to e$ conversion is defined by normalizing to the muon capture rate, $\BR(\mnen) \equiv \Gamma_{\rm conv} / \Gamma_{\rm capt}$. We use the recent
lattice simulation result \cite{Ohki:2008ff} for the $\sigma$ term,
which shows that the strange-quark content of the nucleon is much
smaller than previously thought. Notice that the branching ratio for
$\mu$-$e$ conversion in nuclei is scaled by $\tan^6\beta$, while that
for $\meg$ is proportional to $\tan^2\beta$.

\subsection{Correlation between $\meg$ and $\mnen$}

The ratio of branching ratios $\BR(\mnen)/\BR(\meg)$ is sensitive to new physics parameters. 
In the SUSY scenario, it depends on whether the gaugino-mediated or
Higgs-mediated contribution is dominant. In order to see this, we show the parameter dependence of branching ratios of the cLFV processes.

Assuming only the gaugino- or Higgs-mediated contribution exists, and only the left-handed flavor violation $\delta^{LL}$,\footnote{In the case assuming only the $\delta^{RR}$, Higgs-mediated LFV vanishes for $\mu = \tilde{m}_{L}$ and we cannot use degenerate SUSY mass approximation.} branching ratios of $\meg$ and $\maleal$ processes roughly lead to
\begin{eqnarray}
\BR(\meg)^{L}_{\rm gaugino} \!\! &\simeq& \!\! 4.49 \times 10^{-11} \left({ 25 [\TeV] \over \MSUSY }\right)^{4} \left({ \tb \over 50 }\right)^{2} (\delta^{LL}_{\mu e})^{2}, 
\label{meggauge} \\
\BR(\maleal)^{L}_{\rm gaugino} \!\! &\simeq& \!\! 1.14 \times 10^{-13} \left({ 25 [\TeV] \over \MSUSY }\right)^{4} \left({ \tb \over 50 }\right)^{2} (\delta^{LL}_{\mu e})^{2}, 
\label{mnengauge} \\
\BR(\meg)^{L}_{\rm Higgs} \!\! &\simeq& \!\! 6.14 \times 10^{-12} \left({ 500 [\GeV] \over m_{A^{0}} } \log \big( {m_{A^{0}}^{2} \over m_{t}^{2} } \big) \right)^{4}  \left({ \tb \over 50 }\right)^{2} (\delta^{LL}_{\mu e})^{2}, 
\label{megHiggs}  \\
\BR(\maleal)^{L}_{\rm Higgs} \!\! &\simeq& \!\!1.26 \times 10^{-10} \left({ 500 [\GeV] \over m_{A^{0}}}\right)^{4} \left({ \tb \over 50 }\right)^{6} (\delta^{LL}_{\mu e})^{2}.
\label{mnenHiggs} 
\end{eqnarray}
Here we assume all SUSY particles are degenerate, with a common mass $\MSUSY$ at weak scale. The only top Barr--Zee diagram contribution is written in Eq.~(\ref{megHiggs}) to grasp main $m_{A^{0}}$ dependence: however, it is somehow underestimated.

Since these branching ratios are suppressed by new physics mass scale $\MSUSY$ or $m_{A^{0}}$, the ratio of them is independent to $\delta^{LL}$ or $\delta^{RR}$, and roughly characterized by $\tb$, mass ratio $\MSUSY / m_{A^{0}}$.
If SUSY particles are relatively light, the dipole operator overwhelms both $\meg$ and $\mnen$, and the ratio of the branching ratios becomes constant $\simeq O(\alpha_{\rm em} / \pi)$ \cite{Kitano:2002mt},
\begin{equation}
{\BR(\maleal)^{(L,R)}_{\rm gaugino} \over \BR(\meg)^{(L,R)}_{\rm gaugino}} \simeq { D^{2} G_{F}^{5} \over 192\pi^{2} \Gamma^{\rm Al}_{\rm capt}} \simeq 0.0025,
\label{doublegauge}
\end{equation}
where $\Gamma^{\rm Al}_{\rm capt}\simeq 0.7054 \times 10^{6}
\text{sec}^{-1}$. This value has no information of SUSY parameters.

On the other hand, if SUSY particles are sufficiently heavy, non-decoupling Higgs-contribution becomes dominant. In this case, the ratio is roughly calculated
\begin{equation}
{\BR(\maleal)^{L}_{\rm Higgs} \over \BR(\meg)^{L}_{\rm Higgs}} \simeq {(4\pi)^{4} \over \alpha_{\rm em}^{2}} {2 m_{\mu}^{2} (m_{p} \tilde{C}^{(p)}_{SR} S^{(p)} + m_{n} \tilde{C}^{(n)}_{SR} S^{(n)} )^{2} \over D^{2} N_{c}^{2} Q_{t}^{4} m_{t}^{4}}   \left( \tb \over \log({ m_{t}^{2} \over m_{A^{0}}^{2} }) \right)^{4}.
\end{equation}
It numerically leads to 
\begin{equation}
{\BR(\maleal)^{L}_{\rm Higgs} \over \BR(\meg)^{L}_{\rm Higgs}} \simeq \left( \tb \over 20 \log({ m_{t} \over m_{A^{0}} }) \right)^{4},
\label{doubleHiggs}
\end{equation}
and emerges sensitivity to $\tb$ . 
This sensitivity mainly comes from the tree-level Higgs boson exchange process for $\mnen$, and Barr--Zee diagrams including the bottom-quark loop also contribute secondary. 

Finally, if SUSY particles have moderate weight, there is interference between gaugino- and Higgs-mediated contributions. Then the ratio acquires sensitivity of  mass ratio $\MSUSY / m_{A^{0}}$, 
and in principle, SUSY particle masses. 
To quantify the size of interference, we compare the Higgs- and gaugino-mediated contributions in the same process.

Including parameter dependence of Eq.~(\ref{meggauge}) and Eq.~(\ref{megHiggs}), we estimate the ratio of Higgs- and gaugino-mediated contributions to the $\meg$ process from numerical calculation of the public code that is mainly used in section 3. This value is expressed by
\begin{equation}
 {\BR(\meg)^{(L,R)}_{\rm Higgs} \over \BR(\meg)^{(L,R)}_{\rm gaugino} } = {|A^{(L,R)}_{\rm Higgs}|^{2} \over |A^{(L,R)}_{\rm gaugino}|^{2}} \simeq \left( {\MSUSY \over 55 m_{A^{0}}}     \log \big( {m_{A^{0}} \over m_{t} } \big)   \right)^{4}.
\label{megratio}
\end{equation}
As a result, the Higgs- and gaugino-mediated dipole amplitudes are comparable to each other when $\MSUSY / m_{A^{0}} \simeq 55 / \log ({m_{A^{0}} \over m_{t}  })$ for the $\delta^{LL}$ and $ \delta^{RR}$, both chiralities of flavor-mixing slepton mass terms\footnote{Using the fact that $\log (m_{A^{0}} / m_{t}) \simeq 1$ for $m_{A^{0}}$ = 500[$\GeV$]. }.

The ratio for the $\mnen$ process is estimated similarly as
\begin{equation}
{\BR(\maleal)_{\rm Higgs}^{(L,R)} \over \BR(\maleal)^{(L,R)}_{\rm gaugino} } \simeq \left( {\MSUSY \over 8 m_{A^{0}}} {\tb \over 50} \right)^{4}.
\label{malealratio}
\end{equation}
Then in the region where ${\MSUSY / 8 m_{A^{0}}} \simeq {50 / \tb}$, the Higgs and gaugino-mediated contributions become comparable in the $\mu$ - $e$ conversion process.
The $m_{A^{0}}$ dependence is weaker than the previous ratio of dipole amplitudes, because the dominant Higgs contribution to the $\mu$ - $e$ conversion process is a tree-level Higgs boson exchange and not enhanced by the log factor.

Around the region where these ratios are unity, interference between the Higgs- and gaugino-mediated contributions influences the $\doubleratio$. Consequently, the ratio transits from Eq.~(\ref{doublegauge}) to Eq.~(\ref{doubleHiggs}) and acquires mass parameter dependence. This is what we discuss in the next section.

\section{Mass Spectrum Dependence of $\mu$ - $e$ Transitions and their Correlation in the NUHM Model}

Here we will compare the strength of gaugino- and Higgs-mediated contributions 
and determine the mass-sensitive region of $\BR(\meg) / \BR(\maleal)$,
 in main mass parameter regions of the NUHM, $m_{0}, M_{1/2}, \mu,$ and $m_{A^{0}}$.

\subsection{The non-universal Higgs mass (NUHM) model}
In the MSSM, mSUGRA model 
is an economical framework to describe SUSY phenomenology. 
This model has five free parameters:
\begin{equation}
m_{0}, ~ M_{1/2}, ~ A_{0}, ~ \text{sign} (\mu), ~ \tb,
\end{equation}
and we set these parameters at some high scale.
However, in the mSUGRA model, soft sfermion masses $m^{2}_{\tilde{f}}$ and soft Higgs masses $m^{2}_{H_{1}} ,  m^{2}_{H_{2}}$ 
are equal as $m^{2}_{\tilde{f}} = m^{2}_{H_{1}} = m^{2}_{H_{2}}$ 
and we have no free Higgs mass parameter. 
To analyze the Higgs-mediated contribution in the complete SUSY framework, 
we consider the non-universal Higgs mass (NUHM) model, where 
$m_{H_{1}}, m_{H_{2}}$ are different 
from universal sfermion mass $m_{0}$ at initial scale.
\begin{equation}
m_{0}, ~ m^{2}_{H_{1}}, ~ m^{2}_{H_{2}}, ~ M_{1/2}, ~ A_{0}, ~ \text{sign} (\mu), ~ \tb.
\end{equation}

Using the conditions of electroweak symmetry breaking (EWSB), the initial scale masses $m_{H_{1}}^{2}$ and $m_{H_{2}}^{2}$ are replaced by the weak scale parameters $\mu$ and $m_{A^{0}}$. The tree-level minimization condition for EWSB in the MSSM is given by
\begin{equation}
 \mu^{2} = - {m_{Z}^{2} \over 2} + {m_{H_{1}}^{2} - m_{H_{2}}^{2} \tbb  \over \tbb -1 } ,
 \label{mutree}
\end{equation}
where $m_{Z}$ is the Z boson mass. In addition, the tree-level CP-odd Higgs mass is
\begin{equation}
m_{A^{0}}^{2} = m_{H_{1}}^{2} + m_{H_{2}}^{2} + 2\mu^{2} .
 \label{matree}
\end{equation}
After all, this model has the following free six continuous parameters;
\begin{equation}
 m_{0}, ~ M_{1/2}, ~ A_{0}, ~ \mu, ~ m_{A^{0}}, ~ \tb,
\end{equation}
and permits Higgs and Higgsino masses as free parameters. 

These parameters including $\mu, m_{A^{0}}$ influence the SUSY particle mass spectrum \cite{Ellis:2002wv,Baer:2005bu}.
In particular, the renormalization group running of the soft mass is in general modified from the mSUGRA by the presence of a non-zero $S$ term, contributed by the $U(1)_{Y} ~D$ term
\begin{equation}
 S = m_{H_{2}}^{2} - m_{H_{1}}^{2} + \text{Tr} [m_{\tilde{q}}^{2} - m_{\tilde{l}}^{2} - 2m_{\tilde{u}}^{2} + m_{\tilde{d}}^{2} + m_{\tilde{e}}^{2}   ] = \text{Tr} [ Y_{i} m_{\phi_{i}}^{2} ],
\end{equation}
\begin{equation}
 {dm_{\phi_{i}}^{2}  \over dt} \bigg|_{\rm NUHM}  = {dm_{\phi_{i}}^{2}  \over dt} \bigg|_{\rm mSUGRA} + {1\over16\pi^{2}} {6\over5} g_{1}^{2} Y_{i} S,
\end{equation}
where $t=\log(\mu / \Lambda)$, and $Y_{i}$ is the hypercharge of scalar field $\phi_{i}$.
There are opposite contributions to soft masses following the sign of hypercharge. For example, for large positive $S$, $\tilde{l}_{R}$ are the most suppressed, and sometimes one of them becomes the lightest SUSY particle (LSP), while those of $\tilde{t}_{R}, \tilde{l}_{L}$ are enhanced. And for large negative $S$, vice versa \cite{Baer:2005bu}.
From Eqs.~(\ref{mutree}), (\ref{matree}), Higgs sector parameters at large $\tb$ are approximated as 
\begin{equation}
\mu^{2} \simeq -m_{H_{2}}^{2} |_{m_{Z}}, ~~~ m_{A^{0}}^{2} \simeq m_{H_{1}}^{2} - m_{H_{2}}^{2}|_{m_{Z}},
\end{equation}
and generally large Higgs sector parameters give negative contribution to the $S$: however, this feature is subdominant when $m_{0}, M_{1/2}$ are much larger than $m_{A^{0}}$ and $\mu$.

After this, we present three scenarios. At first, $\mu$ is taken to be the same order to $m_{0}, M_{1/2}$, such as the mSUGRA. In this scenario, non-holomorphic effects do not decouple in the large SUSY particle mass and we call this the {\it non-decoupling scenario}. Second, the Higgsino mass $\mu$ is fixed at some the TeV scale. In this case, non-holomorphic flavor-violating effects do {\it decouple}, and we call this the {\it decoupling scenario}. 
As you will see later, when sleptons are much heavier than charginos (in this scenario, charged Higgsinos), gaugino-mediated dipole amplitudes receive additional log enhancement, in the only left-handed sleptons mixing case. 
As a consequence, $\doubleratio$ approaches a more gaugino-dominant value.
And last, we show the {\it split SUSY scenario} \cite{ArkaniHamed:2004fb}, where the Higgsinos and gauginos are in the around the TeV scale, much lighter than sfermions. 

Throughout this work, we use {\tt suspect} to generate SUSY mass spectra in the numerical analysis   \cite{Djouadi:2002ze}. This code treats $\mu, m_{A^{0}}$, and $\tb$ as weak scale input parameters, while the remaining parameters are GUT scale values. Finally we mention the parameter selection. There are less consistent vacua in the negative Higgsino mass $\mu < 0$ scenario than $\mu > 0$, so we set  $\mu > 0$. In the $\mu < 0$ scenario, numerical behavior is similar.
Since we are interested in the Higgs dominant region, $m_{A^{0}}$ is taken to be around the TeV 
and also assume $A_0 = 0$ for simplicity.

\subsection{Non-decoupling scenario}

Now, we present the {\it non-decoupling scenario}, which has a larger $\mu$ value than the TeV and
the non-holomorphic correction does not decouple when $\MSUSY \to \infty$.
In the mSUGRA, the $\mu$ parameter is approximated from Eq.~(\ref{mutree}):
\begin{equation}
 |\mu|^{2} = - {m_{Z}^{2}\over2} + {1+0.5\tbb \over \tbb -1}m_{0}^{2} + {1+ 3.5\tbb\over\tbb-1}M_{1/2}^{2}, \\
\end{equation}
where the large gaugino mass contribution comes from the gluino loop.
As a  reference value, we take the $\mu$ parameter in the large $\tb$ region
\begin{equation}
\mu = + \sqrt{ 0.5 m_{0}^{2} + 3.5 M_{1/2}^{2} } ~. 
\end{equation}

At first, so as to consider rough behavior of mass dependence, we take universal sfermion and gaugino mass at GUT scale,
\begin{equation}
m_{0} = M_{1/2} = \MSUSY^{\rm GUT}.
\end{equation}
The upper script ``GUT'' is to distinguish the low energy SUSY mass scale $\MSUSY$, often taken to be around the TeV scale. For non-colored SUSY particles participating in this LFV analysis, particle masses at low energy are roughly equal to $ (0.5-2) \MSUSY^{\rm GUT}$.
In this case, the ratio of the branching ratio ${\BR(\maleal) / \BR(\meg)}$ has $\tb$ and mass ratio $\MSUSY / m_{A^{0}}$ sensitivity.

\begin{figure}[h]
\begin{center}
      \includegraphics[width=16cm,clip]{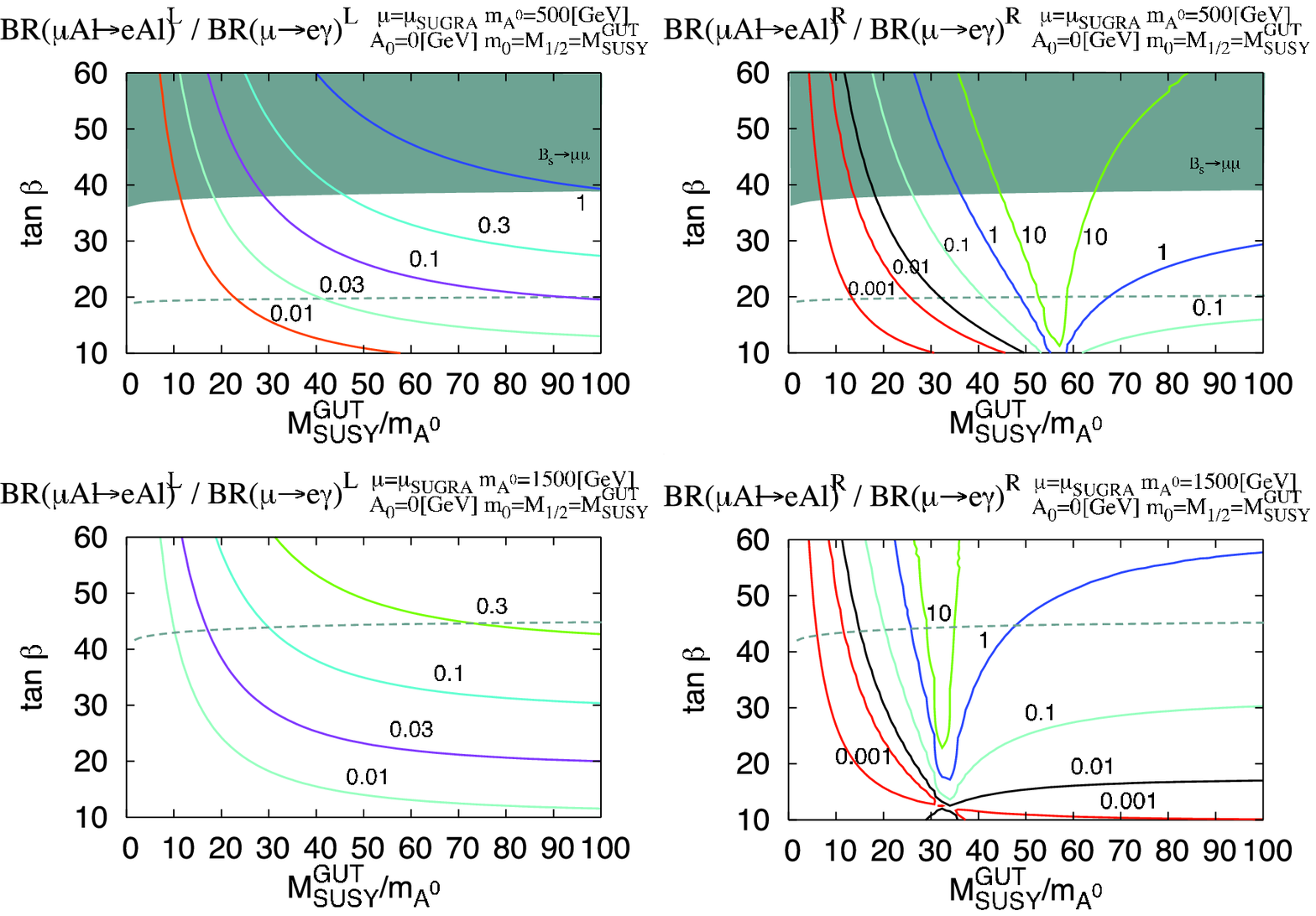} 
   \caption{Contour plot of BR($\mu {\rm Al} \to e {\rm Al}$)
     / BR($\meg$), $\tb$ vs $\MSUSY^{\rm GUT}  / m_{A^{0}}$ including both the
     Higgs- and gaugino-mediated contributions. $m_{A^{0}}$ = 500[\GeV] (upper panels), 1500[\GeV] (lower panels), $\mu = + ( 0.5 m_{0}^{2} + 3.5 M_{1/2}^{2} )^{1/2}$[\GeV] as written in the figures. The dark shaded regions and the dashed lines are the excluded regions and the lines that $\BR(\bsmumu) = 4.3\times10^{-8}, ~ 1.0\times 10^{-9}$, respectively. }
\label{obsvequivtb}
\end{center}
\end{figure}

Fig.~\ref{obsvequivtb} is a contour plot of BR($\mu {\rm Al} \to e {\rm Al}$)
     / BR($\meg$), $\tb$ vs $\MSUSY^{\rm GUT}  / m_{A^{0}}$. We took $m_{A^{0}}$ = 500[\GeV] (upper), 1500[\GeV] (lower), and $\mu = + ( 0.5 m_{0}^{2} + 3.5 M_{1/2}^{2} )^{1/2}$[\GeV] as written in the figures. The dashed-line and gray (dark shaded) region are the line and the bound $\BR(\bsmumu) = 1.0\times 10^{-9}, ~4.3\times10^{-8}$, respectively.

This figure shows that the behavior of this ratio is significantly different for each chirality of flavor violation. 
In the $\delta^{LL}$ case,  there is constructive interference 
between Higgs- and gaugino-mediated contributions, and this ratio increases monotonously for $\MSUSY^{\rm GUT}$.
In the $\delta^{RR}$ case, on the other hand,  there is destructive interference
and strong cancellation for both $\meg$ (the region surrounding the line on which value is 10) and $\mnen$ (the region surrounding the line on which value is 0.001). In this case, it seems to be difficult to give constraints or extract information of parameters from this ratio: however, in some region there  is significant deviation from the standard SUSY prediction, $\doubleratio \simeq O(\alpha_{\rm em} / \pi)$.

In the region where both Higgs- and gaugino-mediated contributions are significant, the $\doubleratio$ becomes sensitive to $\MSUSY^{\rm GUT}/m_{A^{0}}$.
We define this {\it mass-sensitive region}, where the ratios of the Higgs- and gaugino-mediated contributions to $\mnen$ are equal to ${1/ 2}^{4}$ and to $\meg$ are equal to $2^{4}$ (assuming the Higgs-mediated effect is more dominant $\mnen$ than $\meg$). It corresponds to 25\% to 100\% interference at the amplitude level.
From Eq.~(\ref{megratio}) and Eq.~(\ref{malealratio}), we obtain
\begin{equation}
 {\MSUSY^{\rm GUT} \over m_{A^0}} = {200 \over \tb } \sim {110 \over \log ({m_{A^{0}} / m_{t}  })}.
 \end{equation}

This condition corresponds to  $\MSUSY^{\rm GUT} /m_{A^0}\sim (10-110)$ for $m_{A^{0}}=500$ (upper), and  $\MSUSY^{\rm GUT} /m_{A^0}\sim (10-55)$ for $m_{A^{0}}=1500$ (lower). Larger $m_{A^{0}}$ and $\tb$ lead to a  small interference region. 
In this region, both Higgs- and gaugino-mediated diagrams contribute to those
processes in a different way and we could give constraints $\MSUSY^{\rm GUT} /
m_{A^{0}}$ and $\tan\beta$ from $\BR(\mu {\rm Al} \rightarrow e {\rm
  Al}) / \BR(\meg)$.

This behavior is also independent from sign of the $\mu$ parameter, since both Higgs- and gaugino-mediated amplitudes are proportional to $\mu$ (and we assume gaugino masses have the same phase).
In this scenario, the $\bsmumu$ process by non-holomorphic effects does not decouple for large $\MSUSY^{\rm GUT}$, its bound is very strict in the large $\tb$ region when $m_{A^{0}}$ is light \cite{Ellis:2006jy}. 

\begin{figure}[h]
\begin{center}
      \includegraphics[width=16cm,clip]{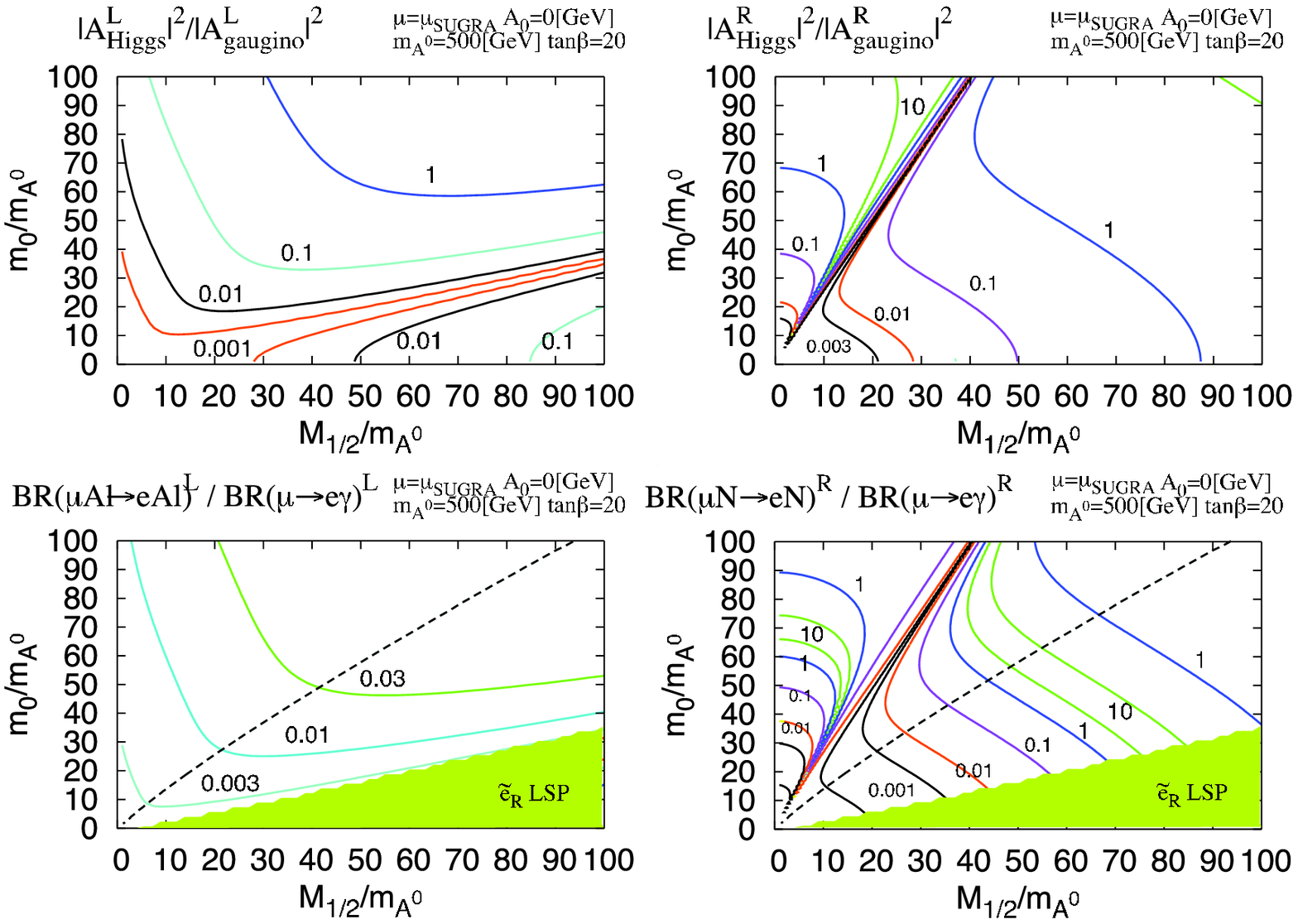}
   \caption{Upper penels : Contour plots of ratio for Higgs- and
  gaugino-mediated contributions to $|A^{(L,R)}|^{2}$ as functions of $\tb, m_0
 /m_{A^{0}}$. Lower panels : Contour plots of BR($\mu {\rm Al} \rightarrow e {\rm Al}$)
     / BR($\meg$), as a function of $\tb$ vs $m_{0} / m_{A^{0}}$. Parameters are taken as written in the figures. The dashed lines are the line that $\BR(\bsmumu) = 1.0\times 10^{-9}$, the value becomes larger in smaller $m_{0}$. In the green (light shaded) regions, $\tilde{e}_{R}$ is the LSP. }
\label{cmptequiv}
\end{center}
\end{figure}

Next, in order to consider detailed cancellation structure, we mention the dependence of the SUSY mass parameters, $m_{0}$ and $M_{1/2}$. 
The upper part of Fig.~\ref{cmptequiv} is contour plots of ratio for Higgs- and 
gaugino-mediated contributions to $|A^{L,R}|^{2}$ as a functions of 
$m_{0}/m_{A^{0}}, M_{1/2} /m_{A^{0}}$. Parameters are taken as written in the figures.
In the $\delta^{LL}$ case, there is strong cancellation in the small $m_{0}$ region. 
This is caused by $\Delta^{L}$, a destructive interference between $SU(2)$ and $U(1)$ contributions in Eq.~(\ref{Delta}). Since we take the large Higgsino mass $\mu = + ( 0.5 m_{0}^{2} + 3.5 M_{1/2}^{2} )^{1/2}$, the right-handed selectron becomes lighter ($\tilde{m}_{e_{R}} \ll \mu$) and $U(1)$ pure $\tilde{B}$ amplitudes overcome $SU(2)$ in the small $m_{0}$ region \cite{Brignole:2003iv}. Of course, this cancellation depends on the reference value of $\mu$.
On the other hand, in the $\delta^{RR}$ case, there is cancellation in both gaugino- and Higgs-mediated  amplitudes.
In the Higgs-mediated contribution, $\Delta^{R}$ becomes zero at $\mu = \tilde{m}_{L}$, and in gaugino-one, at $\mu \simeq \tilde{m}_{L}$. Both effects are originated from the different sign of the hypercharge between $U(1)$ diagrams. For this reason, this amplitude ratio extremely changes around $\mu \simeq \tilde{m}_{L}$: however, physical cancellation occurs between Higgs- and gaugino-mediated amplitudes, around the line on which this ratio is equal to unity. 

The lower part of Fig.~\ref{cmptequiv} shows contour plots of BR($\mu {\rm Al} \rightarrow e {\rm Al}$)
     / BR($\meg$), $M_{1/2} / m_{A^{0}}$ vs $m_{0} / m_{A^{0}}$ including both the
     Higgs- and gaugino-mediated contributions. Parameters are taken as written in the figures.
In the green (light shaded) region, $\tilde{e}_{R}$ is the LSP.
This behavior is caused by the $U(1)$ $S$ parameter, a characteristic feature of the NUHM. When we take large $\mu$, $m_{0}$, and $M_{1/2}$, the $S$ parameter becomes a large positive value at the high (GUT) scale. The positive $S$ makes positive hypercharge fields lighter and negative ones heavier.

In the $\delta^{RR}$ case,  
the Higgs-mediated contribution becomes dominant in the region $\mu \simeq \tilde{m}_{L}$, and the value of BR($\mu {\rm Al} \rightarrow e {\rm Al}$) / BR($\meg$) changes from $O(\alpha / \pi)$.
Although SUSY particle masses are around the TeV scale,
it is necessary to consider the Higgs-mediated contribution 
in the region where the gaugino-mediated contribution receives destructive interference. 

\subsection{Decoupling scenario}
Next, we consider the {\it decoupling scenario}, in which the $\mu$ parameter is around the TeV scale.
In this scenario, non-holomorphic Yukawa effects indeed {\it decouple}, receiving the factor $\mu/\MSUSY^{\rm GUT}$. As a consequence, Higgs-meditated flavor-violating processes such as $\meg$ or $\bsmumu$ also receive this suppression, and these bounds are loose in the large $m_{0} , M_{1/2}$ region.
Theoretically, $\mu$ is the only supersymmetric parameter in the NUHM and 
it need not relate to the soft breaking mass terms. 
If one imposes approximate Peccei-Quinn symmetry, the Higgs sector parameters $\mu$ and $m_{A^{0}}$ are much smaller (and CP-odd Higgs $A^{0}$ becomes axion) than sfermions and might be around the TeV scale.

Contrary to the previous scenario, the strong cancellation in the gaugino-mediated contribution $A^{R}_{\rm gaugino}$ does not emerge in the large $m_{0}, M_{1/2}$ region. 
The reason is that the destructive interference appears only around the $\mu \simeq \MSUSY$, relatively small mass region (in this scenario, around the TeV scale). 

Here we will show that the gaugino-mediated contribution receives additional log enhancement when sleptons are much heavier than charginos (in this scenario, charged Higgsinos), in the only left-handed sleptons mixing case. 
This feature arises from the difference of the asymptotic behavior between gaugino-mediated charged and neutral loop functions in Eq.~(\ref{loopint}).
In the limit as $x \to 0$ (this limit means there are very lighter fermions than scalars), gaugino-mediated loop integrals behave as
\begin{equation}
 f_{n} (x) \to 1, ~f_{c} (x) \to -\ln x, 
\end{equation}
and $f_{c} (x)$ (it comes from charged wino / Higgsino diagrams) contributes only to the $A_{L}$, 
and the decoupling behavior of the ratio of amplitudes differs from the chirality of flavor violation.

Furthermore, the loop function in the non-holomorphic Yukawa behave as
\begin{equation}
 I_{4}(a=b=c=d=M) \to {1\over 6M^{2}} ,
\end{equation}
and for $M \gg \mu$,
\begin{equation}
 I_{4}(a=b=c=M, d=\mu) \to {1\over 2M^{2}} .
\end{equation}
This behavior leads to the amplitude ratio for $\MSUSY \gg \mu$,
\begin{equation}
  {|\Delta^{L}|^{2} \over |A^{L}_{\rm gaugino}|^{2}} \bigg|_{\mu {\rm ~ decoupling}} 
  \simeq \bigg( {1\over 2 \ln ({\MSUSY^{2} / \mu^{2} })} \bigg)^{2}
  {|\Delta^{L}|^{2} \over |A^{L}_{\rm gaugino}|^{2}} \bigg|_{\mu {\rm ~ non~decoupling}} ,
\end{equation}
\begin{equation}
  {|\Delta^{R}|^{2} \over |A^{R}_{\rm gaugino}|^{2}} \bigg|_{\mu {\rm ~ decoupling}} 
  \simeq {1\over 3^{2} }
  {|\Delta^{R}|^{2} \over |A^{R}_{\rm gaugino}|^{2}} \bigg|_{\mu {\rm ~ non~decoupling}} .
\end{equation}
Including these effects, we estimate the ratio of Higgs- and gaugino-mediated contributions to $\meg$ and $\mnen$ in a similar way of the {\it non-decoupling scenario}:
\begin{eqnarray}
{|A^{L}_{\rm Higgs}|^{2} \over |A^{L}_{\rm gaugino}|^{2}} = {\BR(\meg)_{\rm Higgs}^{L} \over \BR(\meg)_{\rm gaugino}^{L} }&\simeq&  \left( {\MSUSY^{\rm GUT} \over 100 m_{A^{0}}}    { \log \big( {m_{A^{0}} \over m_{t} } \big) \over \sqrt{ \log(\MSUSY^{\rm GUT} / \mu)} } \right)^{4}, \label{megratioL} \\
{\BR(\maleal)_{\rm Higgs}^{L} \over \BR(\maleal)_{\rm gaugino}^{L} } &\simeq& \left( {\MSUSY^{\rm GUT} \over 16 m_{A^{0}}} {\tb \over 50 \sqrt{\log(\MSUSY^{\rm GUT} / \mu)}} \right)^{4}, \label{malealratioL} \\
{|A^{R}_{\rm Higgs}|^{2} \over |A^{R}_{\rm gaugino}|^{2}} = {\BR(\meg)_{\rm Higgs}^{R} \over \BR(\meg)_{\rm gaugino}^{R} } &\simeq& \left( {\MSUSY^{\rm GUT} \over 80 m_{A^{0}}}     \log \big( {m_{A^{0}} \over m_{t} } \big)   \right)^{4} , \label{megratioR} \\
{\BR(\maleal)_{\rm Higgs}^{R} \over \BR(\maleal)_{\rm gaugino}^{R} } &\simeq& \left( {\MSUSY^{\rm GUT} \over 10 m_{A^{0}}} {\tb \over 50} \right)^{4} \label{malealratioR}.
\end{eqnarray}
Since this decoupling difference is caused by the charged fermion propagator, similar behavior is expected when $M_{1/2}$ is much smaller than $m_{0}$, as you see in the last scenario. 

\begin{figure}[h]
\begin{center}
      \includegraphics[width=16cm,clip]{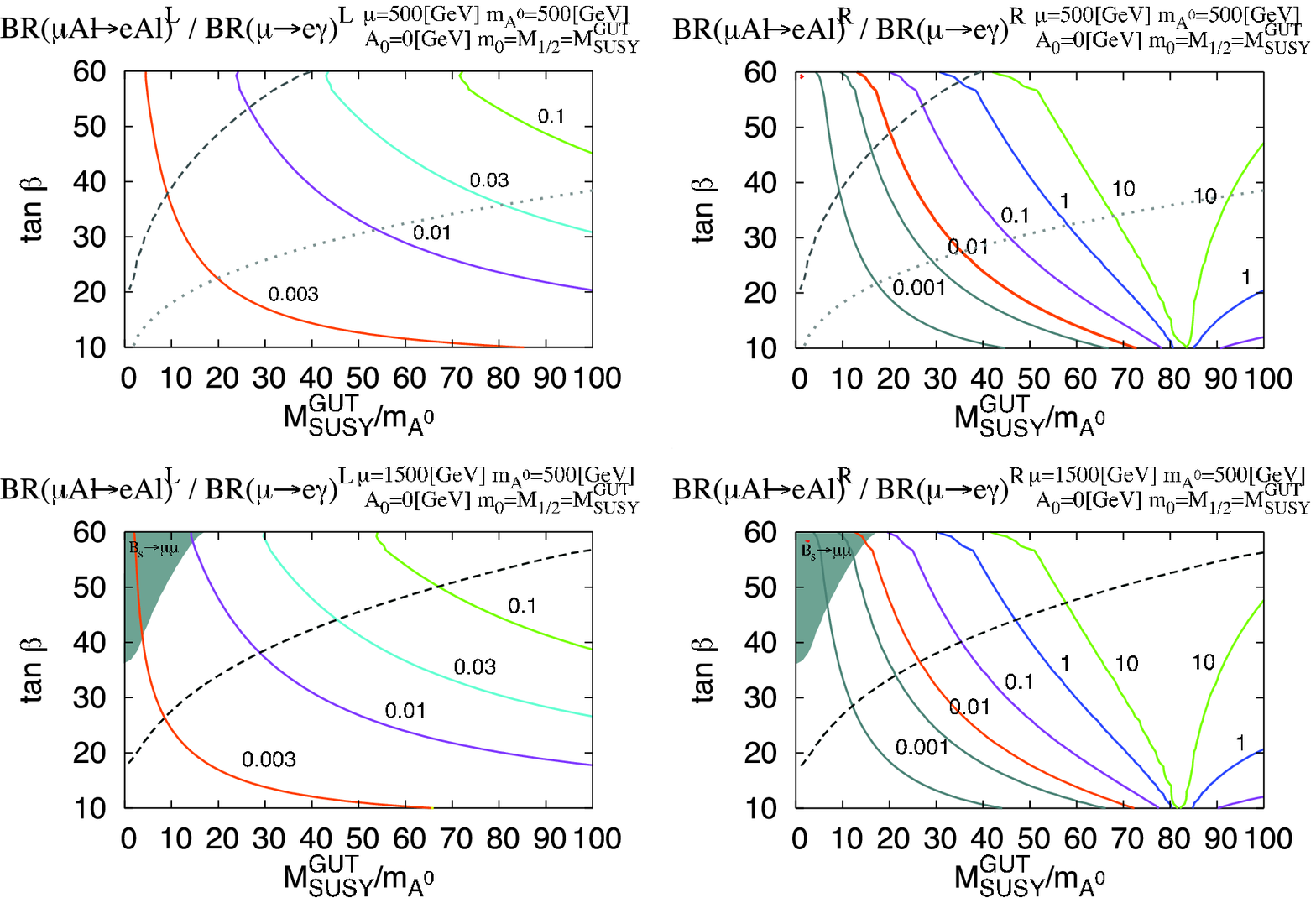}
   \caption{Contour plot of BR($\mu {\rm Al} \rightarrow e {\rm Al}$)
     / BR($\meg$), as function of $\tb$ vs $\MSUSY^{\rm GUT} / m_{A^{0}}$ including both the
     Higgs- and gaugino-mediated contributions. Parameters are taken as written in the figures. The dark shaded regions, the dashed lines and the dotted lines are the excluded region and the line where $\BR(\bsmumu) = 4.3\times10^{-8}, ~ 1.0\times 10^{-9},  ~ 1.0\times 10^{-11} $, respectively.}
\label{obsvtb}
\end{center}
\end{figure}

Fig.~\ref{obsvtb} is a contour plot of BR($\mu {\rm Al} \rightarrow e {\rm Al}$)
     / BR($\meg$), as a function of $\tb$ vs $\MSUSY^{\rm GUT} / m_{A^{0}}$. Parameters are taken as written in the figures. The dashed-lines and gray (dark shaded) regions are the line and the bound $\BR(\bsmumu) = 1.0\times 10^{-9}, ~4.3\times10^{-8}$, respectively.
In the $\delta^{LL}$ case, since there is log enhancement of gaugino contributions,
the dipole amplitude dominates for both $\meg$ and $\maleal$. 
This ratio approaches more a gaugino-dominant value,
 $\BR(\maleal) / \BR(\meg) \simeq O(\alpha_{\rm em} / \pi)$ and $\tb$ dependence becomes weaker than the previous scenario
for the same $\MSUSY^{\rm GUT} / m_{A^{0}}$ value.  
On the other hand, in the $\delta^{RR}$ case, $\mu$ dependence is negligible since $A^{R}$ receive no $\sqrt{\log(\MSUSY^{\rm GUT} / \mu)}$ factor. 
There is cancellation in both $\BR(\meg)$ and $\BR(\maleal)$, similar to the previous scenario.

The mass-sensitive region is defined where the values of Eqs.~(\ref{malealratioL}), (\ref{malealratioR}) are equal to $1/2^{4}$ and (\ref{megratioL}), (\ref{megratioR}) are equal to $2^{4}$. In this scenario, for the $\delta^{LL}$ case, this region is roughly estimated as
\begin{equation}
 {\MSUSY^{\rm GUT} \over m_{A^0}} = 400  {\sqrt{\log(\MSUSY^{\rm GUT} / \mu)} \over \tb } \sim  200 {\sqrt{\log(\MSUSY^{\rm GUT} / \mu)} \over \log ({m_{A^{0}}/ m_{t}  })}  ,
\end{equation}
and for the $\delta^{RR}$ case,
\begin{equation}
 {\MSUSY^{\rm GUT} \over m_{A^0}} = {250 \over \tb } \sim {160 \over \log ({m_{A^{0}} / m_{t}  })}.
 \end{equation}

As a result, we found that the previous result is no longer valid when sleptons are much heavier than charginos, 
 and this ratio drastically depends on the mass spectrum structure and chirality of flavor violation.
This condition corresponds to  $\MSUSY^{\rm GUT} /m_{A^0}\sim 20 - 1050$ for $\mu=500, ~ \delta^{LL}$ (upper left panel), $10 - 450$ for $\mu=1500, ~ \delta^{LL}$ (lower left panel), and  $10-160$ for $\delta^{RR}$ (right panels). 
Larger $m_{A^{0}}$ and $\tb$ lead to a small interference region. 
Interestingly, in this scenario, the log factor $\log (\MSUSY^{\rm GUT}/\mu)$ in the only $\delta^{LL}$ case leads to a very large mass-sensitive region. Physically, the gaugino-mediated contribution decouples more slowly by the log enhancement and interferes with the non-decoupling Higgs effect significantly.
The $\bsmumu$ bound is looser than the {\it non-decoupling scenario}, however, there is some constrained region for large $\mu$ and small $m_{A^{0}}$ value.

\subsection{Split SUSY scenario}

Finally, we will show the {\it split SUSY scenario} \cite{ArkaniHamed:2004fb}, which has relatively lighter gauginos, Higgsinos, and CP-odd Higgs boson than sfermions.

\begin{figure}[h]
\begin{center}
      \includegraphics[width=16cm,clip]{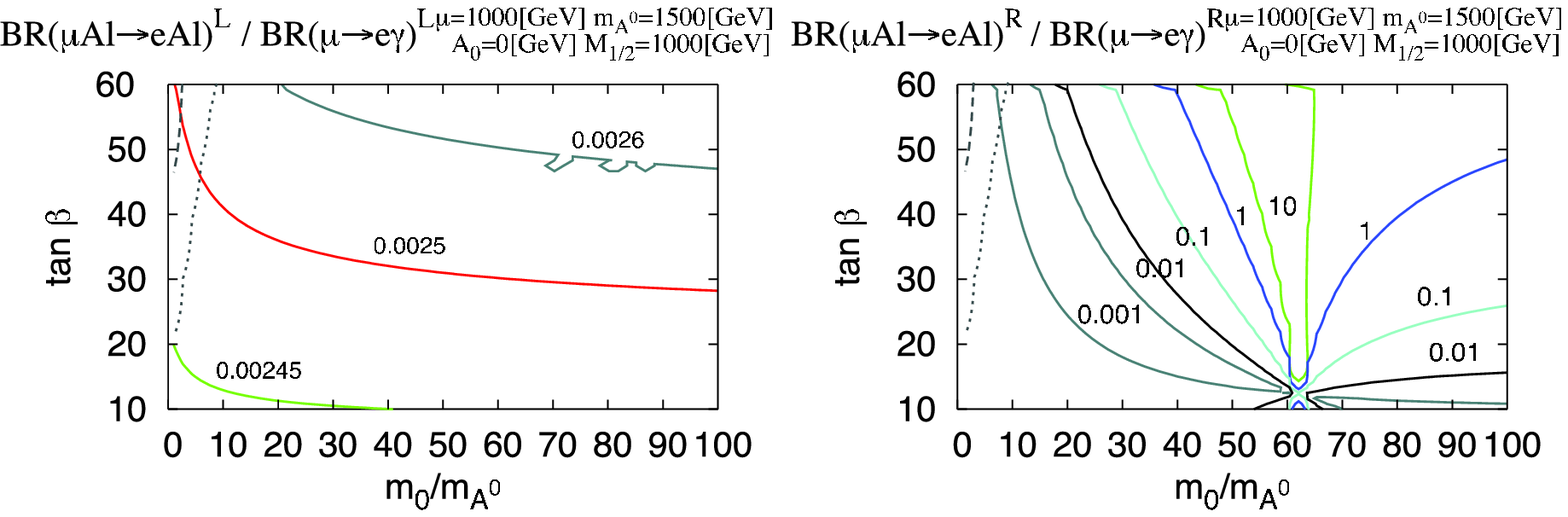} 
   \caption{Contour plots of ratios for Higgs- and
  gaugino-mediated contributions to $|A^{(L,R)}|^{2}$ as functions of $\tb, m_0
 /m_{A^{0}}$. Contour plots of BR($\mu {\rm Al} \rightarrow e {\rm Al}$)
     / BR($\meg$), as functions of $\tb$ vs $m_{0} / m_{A^{0}}$. Parameters are taken as written in the figures. The dashed lines and the dotted lines are the lines where $\BR(\bsmumu) = 1.0\times 10^{-9}, and 1.0\times 10^{-11}$, respectively.}
\label{split}
\end{center}
\end{figure}

 Fig.~\ref{split} shows contour plots of ratio for Higgs- and
  gaugino-mediated contributions to $|A^{(L,R)}|^{2}$ as functions of $\tb, m_0
 /m_{A^{0}}$, and the lower parts of Fig.~\ref{split} are plots of BR($\mu {\rm Al} \rightarrow e {\rm Al}$)
     / BR($\meg$), as a function of $\tb$ vs $m_{0} / m_{A^{0}}$. Parameters are taken as written in the figures.

In the $\delta^{LL}$ case, the log enhancements from two lighter charged fermions, Higgsinos and winos, become more considerable and the Higgs-mediated contribution is negligible in this figure. As a consequence, the $\BR(\maleal) / \BR(\meg)$ is almost constant, $\simeq O(\alpha_{\rm em} / \pi)$. On the other hand, in the $\delta^{RR}$ case, the behavior does not change so much. In this scenario, the mass-sensitive region in the $\delta^{LL}$ becomes more a large $\MSUSY/m_{A^{0}}$ value and more vast. There is a clear difference of the physical observable on the chirality of flavor violation.

\section{Conclusions and Discussion} 

In this paper, we study non-decoupling $\mu$ - $e$ transition effects by a Higgs-mediated contribution in the MSSM, when some SUSY mass parameters are much greater than the TeV scale. In order to treat CP-odd Higgs mass $m_{A^{0}}$ as a free parameter, we consider the non-universal Higgs mass model (NUHM), and assume the only the left- or right-handed sleptons had flavor-mixing mass terms. 

In a previous study \cite{Hisano:2010es}, we have discovered that Higgs- and gaugino-mediated 
dipole amplitudes become comparable to each other when $\MSUSY/m_{A^0} \sim 50$. It was assumed that only left-handed sleptons had flavor-mixing terms and degenerate SUSY particle mass $\MSUSY$ which was much greater than the TeV scale.

However, we found that the previous result is no longer valid with some flavor-violation sources and mass spectrum of the NUHM. 
When only right-handed sleptons have flavor violation, there is some Higgs-dominant region although SUSY particle masses are around the TeV scale.
We found it is necessary to consider the Higgs effect in the region where the gaugino effect receives destructive interference.  
Moreover, when sleptons are much heavier than charginos, cf. split SUSY models,  
we found that the ratio $\BR(\meg) / \BR(\maleal)$ drastically depends on the mass spectrum structure and chirality of flavor violation.
The log factor from two split mass scale influences the way of interference between gaugino- and Higgs-mediated contributions significantly.

Using additional cLFV or chirality information, such as 
 the $\mu$ - $e$ conversion process with different nuclei, 
or P-odd asymmetry of initial polarized $\meg$ decay \cite{Okada:1999zk} ,
we might approach the structure of SUSY mass parameters and the origin of the LFV sources.

\section*{Acknowledgment}
The author is most grateful to J. Hisano for valuable discussions and continuous support 
and also would like to thank S.Sugiyama and M.Yamanaka for their cooperation in the early stages.

{}

\end{document}